\documentclass[12pt, preprint]{aastex}
\usepackage{natbib}
\usepackage{float}
\usepackage{graphicx}
\usepackage{amssymb, amsmath}
\usepackage{hyperref}
\usepackage{multicol}
\usepackage[all]{hypcap}

\title{Isolating the young stellar population in the outer disk of NGC~300\footnotemark \footnotetext{This paper is based on Hubble Space Telescope observations.}}
\author{Tristan J. Hillis\altaffilmark{1}, Benjamin F. Williams\altaffilmark{1}, Andrew E. Dolphin\altaffilmark{2}, Julianne J. Dalcanton\altaffilmark{1}, Evan D. Skillman\altaffilmark{3} }

\altaffiltext{1}{Department of Astronomy, University of Washington, Box 351580, Seattle,
	WA, 98195}
\altaffiltext{2}{Raytheon Company, 1151 E. Hermans Road, Tucson, AZ 85756, USA}
\altaffiltext{3}{Minnesota Institute for Astrophysics, 116 Church St. SE, Minneapolis, MN 55455, USA}

\begin{document}

\begin{abstract}
	The recent star formation history (SFH) in the outer disk of NGC~300 is presented through the analysis of color magnitude diagrams (CMDs).  We analyze resolved stellar photometry by creating CMDs from four \textit{Hubble Space Telescope} fields containing a combination of images from the \textit{Advanced Camera for Surveys} and the \textit{UVIS} imager aboard the \textit{Wide Field Camera 3}.  From the best models of these CMDs, we derive the SFH in order to extract the young stellar component for the past 200 Myrs.  We find that the young stellar disk of NGC~300 is unbroken out to at least $\sim$8 scale lengths (including an upper limit out to $\sim$10 scale lengths) with $r_s=1.4\pm0.1$ kpc, which is similar to the total stellar surface brightness profile.  This unbroken profile suggests that NGC~300 is undisturbed, similar to the isolated disk galaxy NGC~2403.  We compare the environments of NGC~300, NGC~2403, and M33 along with the properties of the gas and stellar disks.  We find that the disturbed HI outer disk morphology is not accompanied by a break in the young stellar disk.  This may indicate that processes which affect the outer HI morphology may not leave an imprint on the young stellar disk.

\end{abstract}

\defcitealias{2013ApJ...765..120W}{W13}

\section{Introduction}
			A fundamental question in galaxy evolution is how the environment transforms disk galaxies.  External processes can have a great effect on galaxies; in galaxy clusters there is a lower number of spiral types compared to field galaxies \citep{1980ApJ...236..351D}, and spirals that are clustered are observed to be redder than their non-clustered counterparts \citep{1983AJ.....88..483K, 1994A&A...289..715D}.  In simulations, spiral galaxies interacting, either through mergers or close orbits, can become elliptical or S0 galaxies \citep{1985A&A...144..115I,1998ApJ...502L.133B}.  Such gravitational interactions all impact the evolution of spiral galaxies.\\
			\indent In particular, the outer disks of spirals give us a fossilized window into their evolution.  Truncations in the exponential profile of spirals, known as disk breaks, may then provide clues about the effects of interactions on these outer disks.  Present studies on disk breaks have built upon the work of \citet{1970ApJ...160..811F}, who began to classify spiral disks into different types, outlined nicely by \citet{2008AJ....135...20E}.  \citet{2008AJ....135...20E} performed an extensive statistical study on spiral galaxies \citep[see also][]{2006A&A...454..759P,2011AJ....142..145G} finding that the majority of late-type spirals have truncations in their exponential disk.  Furthermore, \citet{2012ApJ...744L..11E} and \citet{2012MNRAS.419..669M} find there is no difference between the disk-profiles of cluster galaxies and field galaxies, suggesting global galaxy environment does not play a significant role in the origin of breaks.  Other local effects could then be the main driving force for outer disk evolution, but it is still a mystery why breaks would be independent of global environment.\\	
			\indent Individualized observations of the outer regions of local disks may help constrain current understanding of environmental effects in their evolution.  Fortunately, there are three, representative, nearby disk galaxies in differing environments: M33, NGC~2403, and NGC~300.  These are the closest pure disk galaxies for comparing the effects of different environments on disk evolution \citep[see Figure 1 and Table 1][hereafter W13]{2013ApJ...765..120W}.  M33 is the least isolated, containing a warped HI disk likely due to a weak interaction with M31.  NGC~2403 is the most isolated with no large nearby companion and an undisturbed HI field \citetext{\citetalias{2013ApJ...765..120W}; and references therein}.  NGC~300, the focus of this study and of SA(s)d type \citep{1991rc3..book.....D}, is relatively isolated in that it has no massive companion.  However, it does have a nearby low mass companion, NGC~55, and presents a severe HI warp in its outer parts which may be due to the closeness of this companion \citep{1990AJ....100.1468P}.\\
			\indent The properties of the outskirts of these galaxies are probes of how the structure of outer disks may be related to their environment.  Current observations of NGC~2403 show it to have no break \citepalias{2013ApJ...765..120W} in the young stellar component, while M33 does have a break possibly created from nearby M31 \citep{2007iuse.book..239F}.  NGC~300 has no break in its overall surface brightness profile \citep{2005ApJ...629..239B}, suggesting little environmental influence. Based on M33 and NGC~2403, we might expect that environment affects outer disk structure.  If so, NGC~300 probes an intermediate density environment, where we can obtain detailed information about outer disk structure.  If the young component shows no evidence of a break, then the warp in the gas disk is likely due to environmental effects, as in M33; however, if the young component mimics that of NGC~2403, then environment is less likely to have influenced the structure of the outer disk of NCG~300. \\
			\indent We extend the study of \citetalias{2013ApJ...765..120W} to examine the structure of the young outer disk of NGC~300 following a similar methodology, in this paper, to measure the recent star formation history (SFH) of the resolved stellar populations out to a galactocentric radius of $\sim$14 kpc in order to isolate the star-forming disk from the total stellar profile.  In Section \hyperref[sec:dataAnalysis]{2} we present the data along with the reduction and analysis techniques applied.  Section \hyperref[sec:results]{3} discusses our findings of NGC~300 including a comparison with NGC~2403 and M33, and we finish with a summary in Section \hyperref[sec:conclusion]{4}.\\
			\indent We assume an inclination of $i=42.3^{\circ}$ \citep{1983A&A...118....4B} and a distance modulus of 26.5 from tip of the red giant branch measurements with similar Hubble Space Telescope (HST) fields (2.0 Mpc, \citealt{2009ApJS..183...67D}).\\

	\section{Data \& Analysis}\label{sec:dataAnalysis}
	
	\subsection{Data}
\begin{figure}[b]
		\centering
		\includegraphics[scale=0.5]{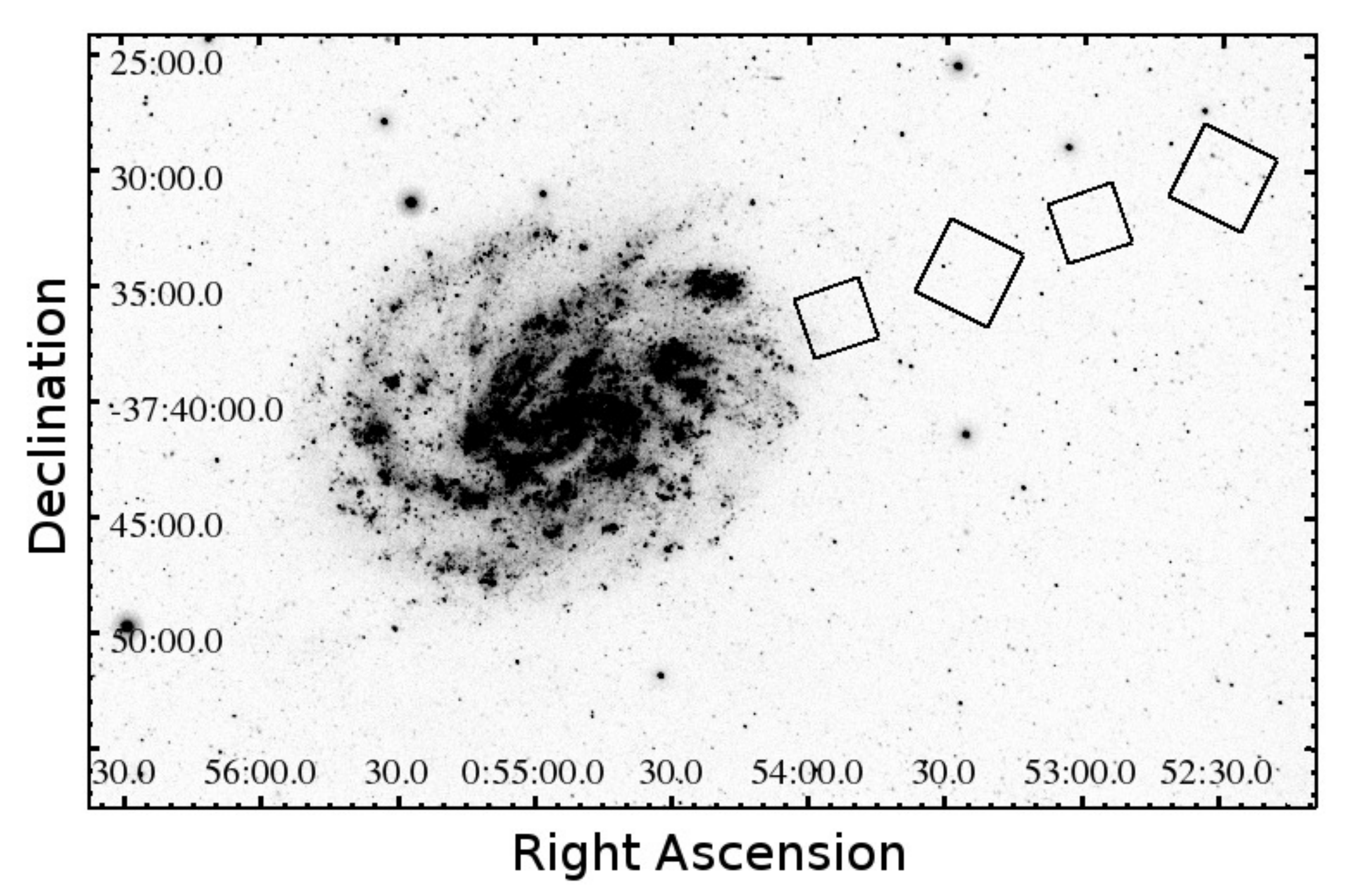}
		\caption{GALEX near-ultraviolet image of NGC~300 with our HST field images overlaid.  Ordering for the fields, from inside out, is field 1, 2, 3, and 4; see Table \ref{tab:fields} for specifics on each field.  Note that our scale is 9.70 pc/arcsec.}
		\label{fig:ngcfield}
	\end{figure}

\begin{figure}[b]
		
		\begin{multicols}{2}
			\includegraphics[scale=0.45]{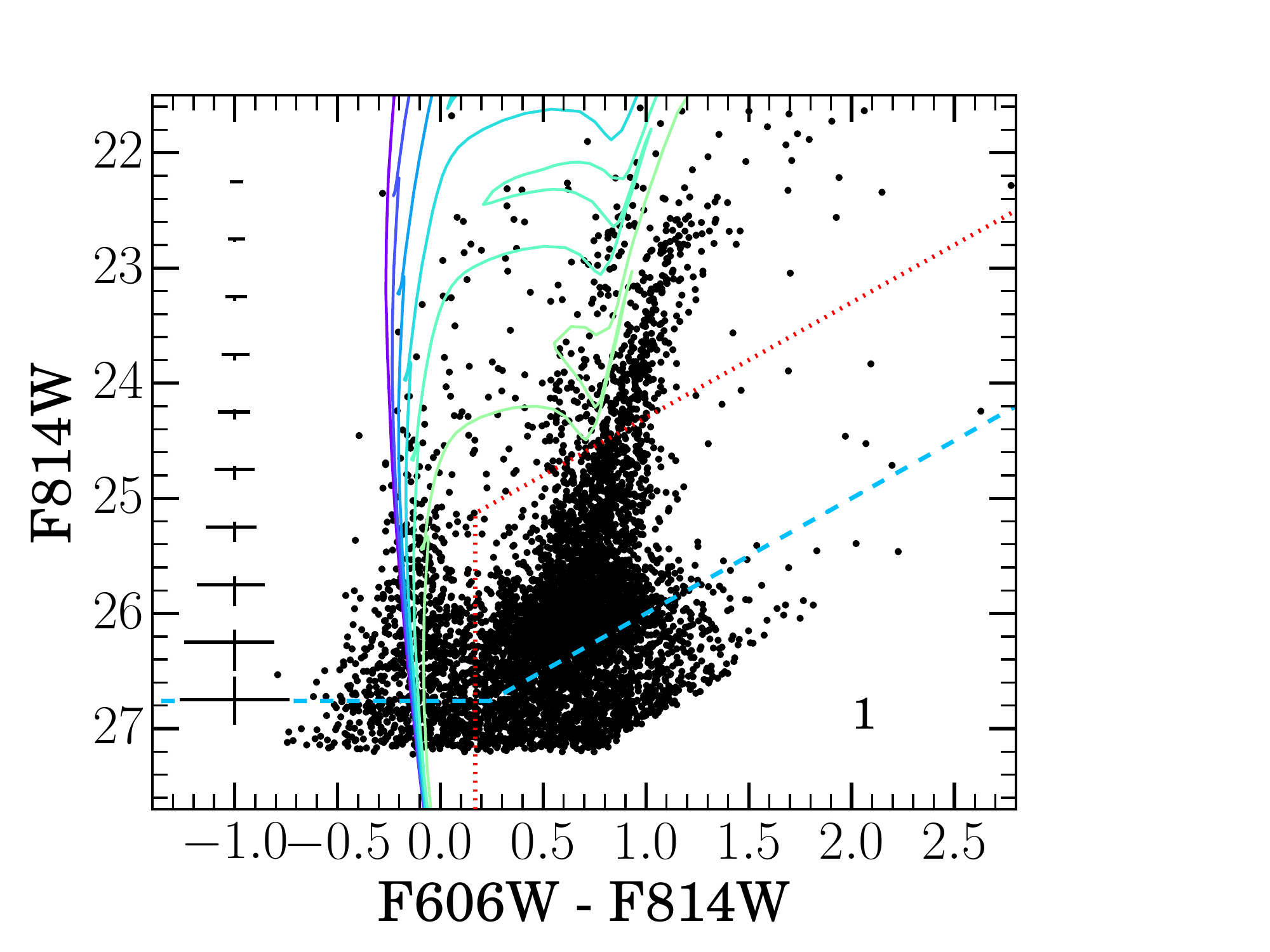}
			\includegraphics[scale=0.45]{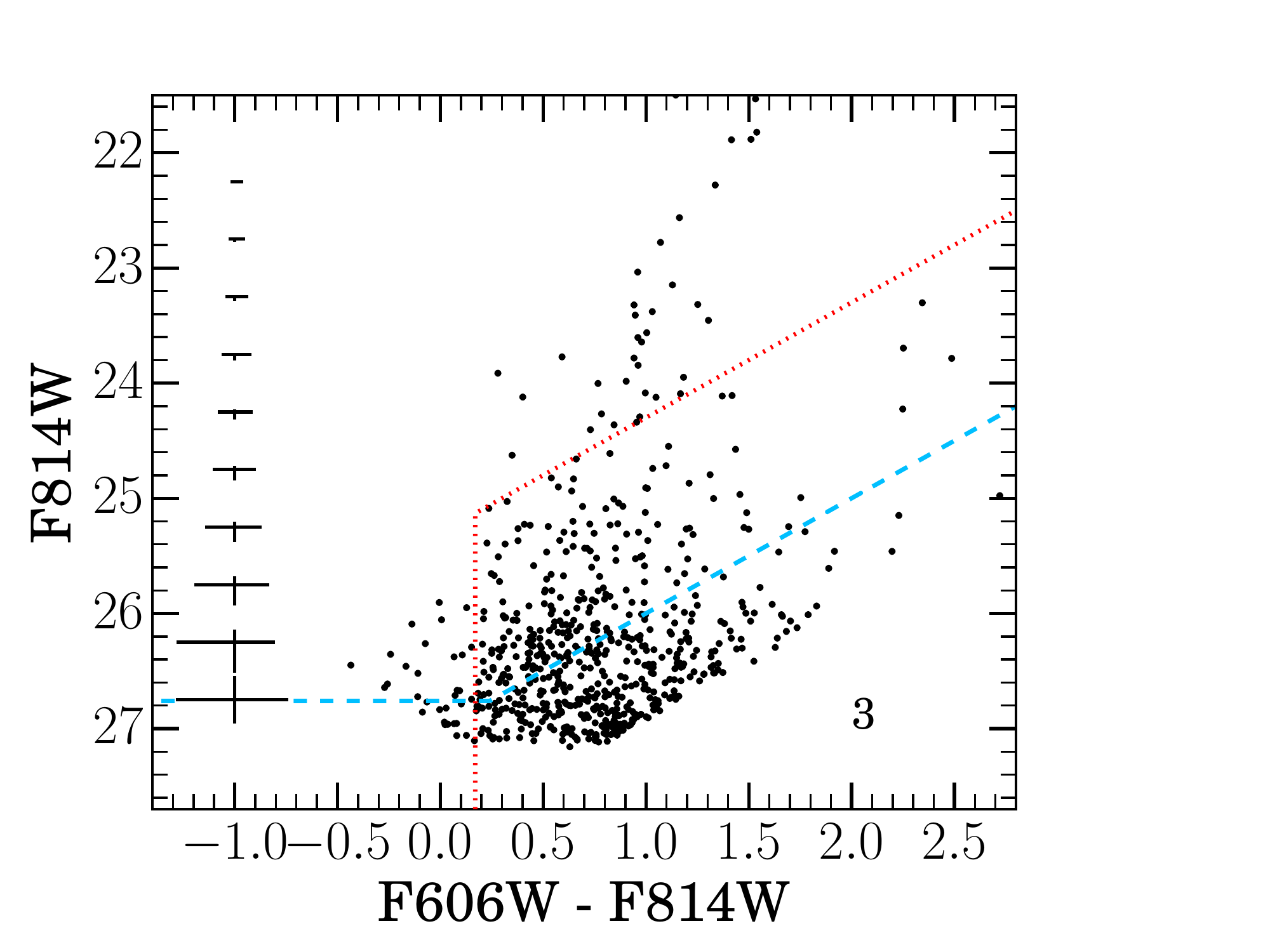}
			\includegraphics[scale=0.45]{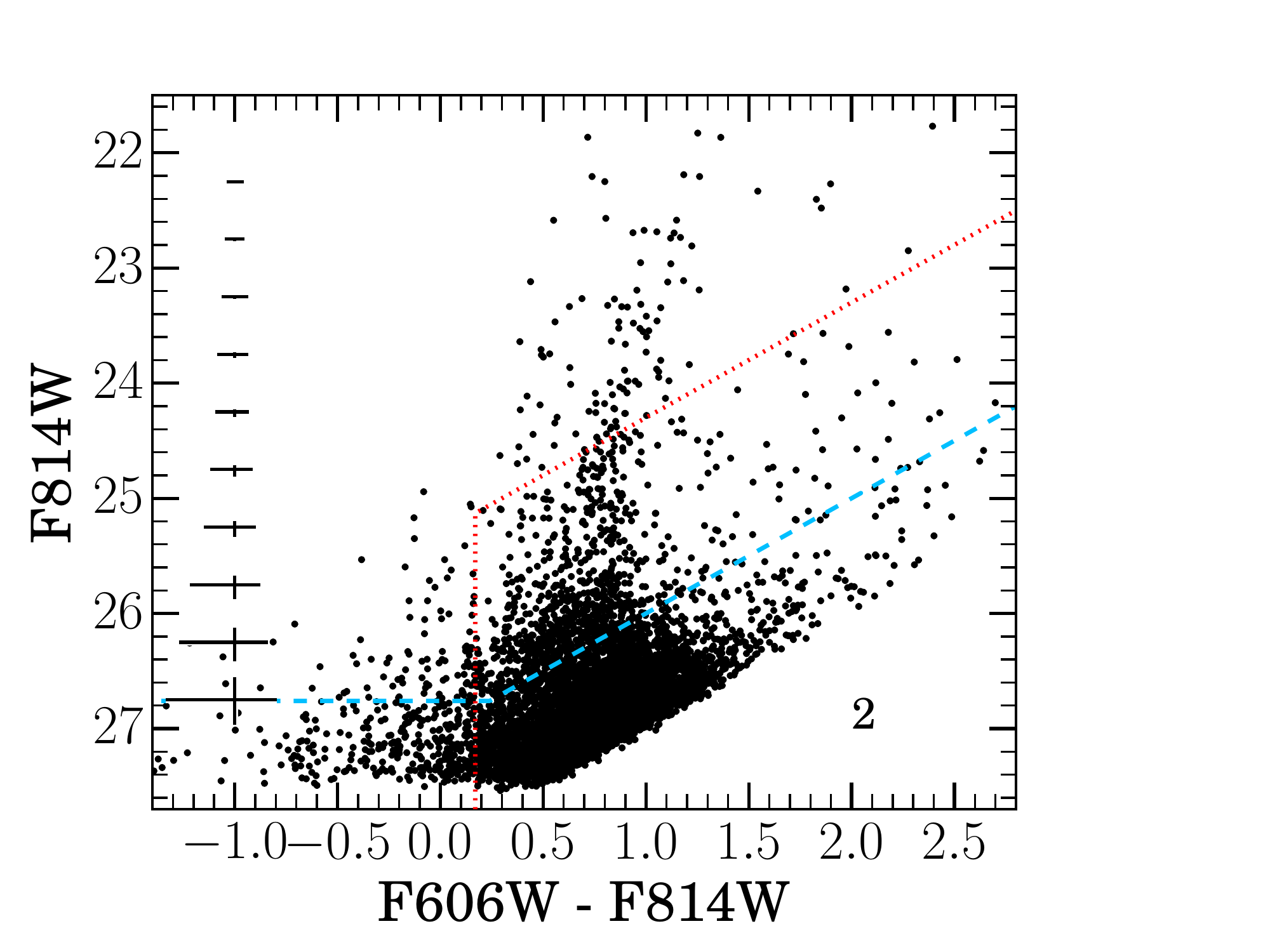}
			\includegraphics[scale=0.45]{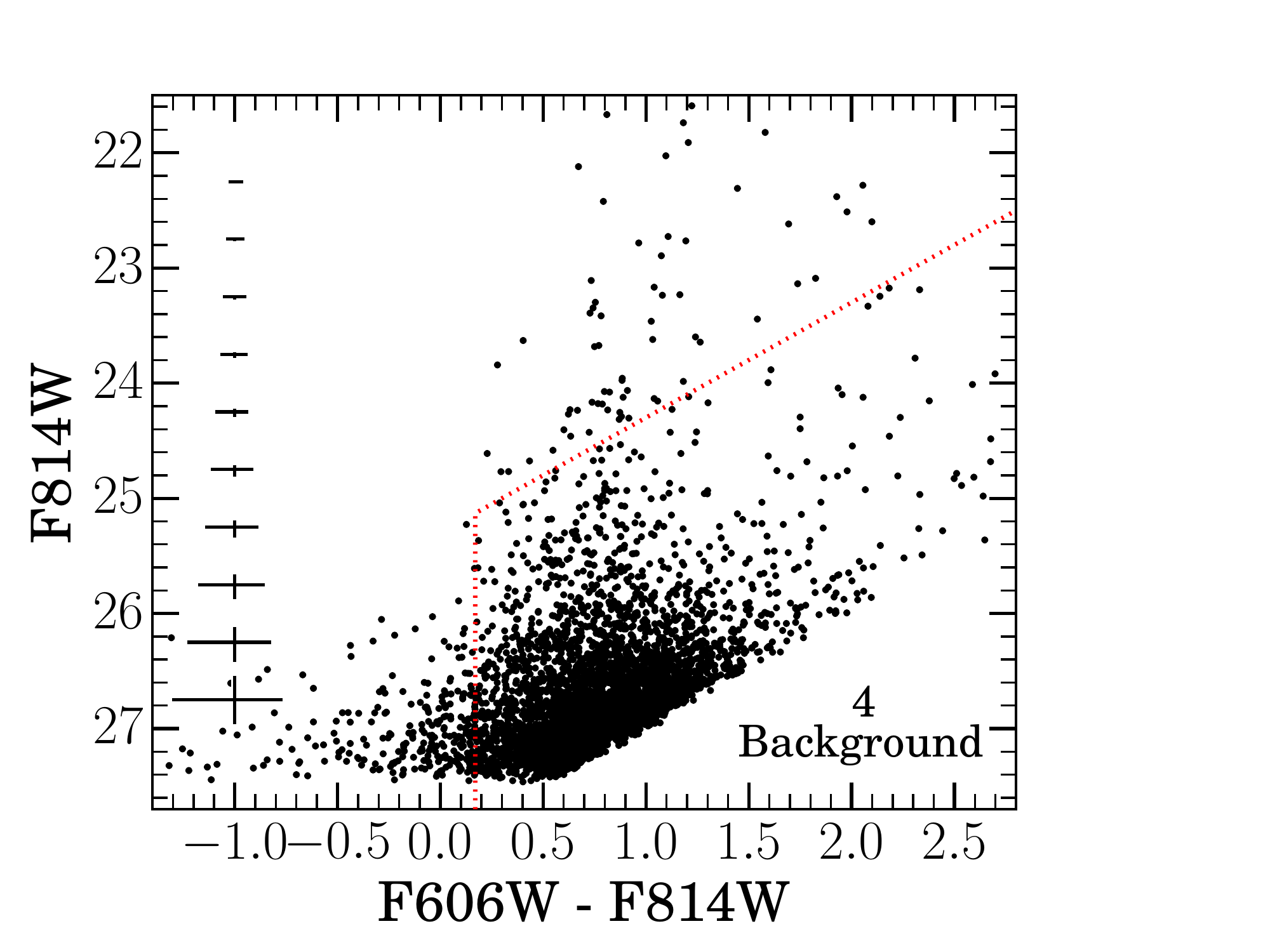}
		\end{multicols}
				
		\caption{CMDs of each HST field given in increasing radius from left to right.  We are looking to model the blue earlier stars, made up of main-sequence and helium burning sequence stars, to find the recent SFR.  Plotted in a blue dashed line are the chosen completeness limits used in the MATCH fitting, and plotted to the left are the fake star recovery errors.  Stars below and to the right of the red dotted outline are excluded during a separate test for robustness in our results (see Section \ref{sec:match}).  In the first frame, are a sample of the isochrone models used; the plotted isochrones span 7 to 8.3 log years.  The last CMD, number 4, helps to model the background detections in the other three fields and is labeled thusly.   Note that we assume a distance modulus of 26.5, or 2.0 Mpc.}
		\label{fig:cmd}
	\end{figure}

		\indent The data presented in this analysis were acquired by HST for GO-13461 (PI: Benjamin Williams).   In Figure \ref{fig:ngcfield} we show a near ultraviolet GALEX image of NGC~300 with each of our HST fields overlaid.  In addition, we present a summary of each image captured in Table \ref{tab:fields}.  The first column outlines the target name and designated field number; each image is listed in ascending order by radius.  Column two and three specify which camera and its associated filter were used with the exposure times given in column four.  We give the galactocentric distance in column five with the area of each field in column six, a conversion factor of 9.7 pc/arcsecond and an inclination correction of $i=42.3^\circ$ were used. \\
		\indent For each of these 4 fields we measured resolved stellar photometry using the photometric pipeline from the Panchromatic Hubble Andromeda Treasury (PHAT). The details of this pipeline are given in \citet{2014ApJS..215....9W}. In short, the pipeline uses the DOLPHOT photometry package, which is an updated version of HSTPHOT \citep{2000PASP..112.1383D}. The individual CCDs from each exposure are prepared by multiplying them by the appropriate pixel area map, and masking bad pixels and cosmic ray hits as determined by the PyRAF task {\it astrodrizzle}. The images are then stacked in memory to search for peaks that may correspond to point sources. Point spread function (PSF) fitting is then forced at all of these locations in all of the exposures. The resulting catalogs contain the magnitude and quality parameters in each exposure and combined for each filter observed. The results are then filtered on the quality	parameters to remove unreliable measurements, as well as many contaminants using a S/N cut at 4.0 and crowding cut at 1.3 and 2.25 for UVIS and ACS imagers, respectively.  Once the photometry was complete, 10$^5$ fake stars covering the relevant range of color and magnitude space were then inserted (one at a time) into the data to determine completeness and uncertainty as a function of color and magnitude in each field.  These artificial stars also undergo identical quality cuts using S/N and crowding including a sharpness cut of 0.15 and 0.2 for the UVIS and ACS imagers, respectively.  The number of observed and fitted stars used in our analysis is given in column seven of Table \ref{tab:fields}, and the 50$\%$ completeness limit, determined from the fake stars, is seen in last column.\\
		\indent The PHAT pipeline is inclusive to crowding stars, optimized for detecting packed stars like in a face-on galactic disk; however our fields are conversely so, and filled with a number of field galaxies being falsely detected.  We applied a stricter cut of the crowding limit to better filter out any galaxies being shredded by the pipeline into false star clusters.  The cut is optimized to exclude as many false detections as possible while maintaining our positive detections.  In addition, with the results being sensitive to a small number of detections we check by eye for false positives due to foreground star diffraction spikes or detector defects.\\
		\indent In Figure \ref{fig:cmd} we present the color magnitude diagrams (CMD) for each field in ascending order of radius from left to right.  There are more star counts in the ACS images because of its larger field of view and higher red sensitivity.  Plotted in the first CMD are a sample of the set of isochrones used in the modeling that range from 7.0 to 8.3 log years.  Plotted in a dashed blue line, is the our defined completeness limit, detections below it are not include in the fitting process (see section \ref{sec:match}).  The last CMD, field 4, is labeled as ``background" because it is used as a measurement of the background contaminants of the first 3 fields (see Section \ref{sec:background}).\\
		\indent Column seven of Table \ref{tab:fields} excludes star counts below the completeness limit shown in Figure \ref{fig:cmd}.  Furthermore, we are only interested in the young population, which represents a further fraction of the perceived counts.  In this way, the data making up the young stellar population ranges from roughly several hundred in field 1 to low double digits in field 3 of which these fields have additional background detections that will be excluded.

	\subsection{Synthetic CMD Fitting}
		\subsubsection{Using MATCH} \label{sec:match}
			\indent  Synthetic CMD fitting of each field, shown in Figure \ref{fig:cmd}, was accomplished using the software package MATCH \citep{2002MNRAS.332...91D} which finds the best linear combination of synthetic CMDs for a specified age and metallicity range.  The best fit corresponds to the SFH output, which is then repeated for each field.  The general process of the use of MATCH is outlined in greater detail in \citet{2009AJ....137..419W}, but we detail the inputs used here.\\
			\indent We assume a Kroupa IMF \citep{2001MNRAS.322..231K} and assume a distance modulus of 26.5.  In addition we assume an extinction of 0.035 \citep{2011ApJ...737..103S} because MATCH is not sensitive to this precise of measurement.  Our SFH fitting is specified with a range of metallicity from -2.3 to 0.1 in dex of 0.1.  Also, we constrain the metallicity enrichment to be constant, or monotonically increasing \citetext{see \citealt{2011ApJ...739....5W} for similar technique}, as the photometry is not deep enough to fully constrain the chemical enrichment history.  With this, we constrain the present day metallicity between -1.0 and -0.5 to reflect more sensitive measurements, such as from planetary nebulae \citep{2013A&A...552A..12S}.  We then specify the time bins with 0.1 dex ranging from 6.6 to 10.1 log years in conjunction with the Padova isochrone set \citep{2008A&A...482..883M, 2010ApJ...724.1030G} that span 6.6 to 10.15 log years with a dex of 0.5. This defines fitting for stars across multiple epochs of which we are most interested in main-sequence and helium burning sequence stars within the first 200~Myrs (6.6 to 8.3 log years).\\
			\indent  We specify an appropriate range of color with a minimum of -0.5 to 3.0 with a dex 0.05.  Along with color, a minimum and maximum magnitude to the corresponding filters is specified with the brightest detected star and the $50\%$ completeness value, in Table \ref{tab:fields}, respectively.  For the ACS field, the 50\% completeness limit (that defines our maximum magnitude) was fainter than the UVIS limit.  We opt to then match the brightest UVIS limit for F814W, while using a value slightly brighter than this limit in F606W as our maximum magnitude across all fits to sufficiently model our CMD and maximize homogeneity while avoiding the less reliable low signal-to-noise measurements.  \\
			\indent We test different MATCH settings to quantify the differences in the chosen procedure.  Invoking the differential extinction flag, ``dAv", we can test the effective of differential extinction in the field dust by allowing us to set maximum differential extinction.  We do not expect large quantities of dust in the outer disk to cause variability in extinction across the field of NGC~300, and unsurprisingly the best fits worsen.  In addition, we measured the star formation history while excluding all stars below and to the right of the red dotted line (drawn from1 (0.17, 27.9) to (0.17, 25.13) and ending on (3.0, 22.3) in Figure \ref{fig:cmd}) .  These results were found to be within uncertainties, demonstrating the robustness of the measured recent young star formation history.

		\subsubsection{Background Contaminates}\label{sec:background}
			At these radii, in the NGC~300 disk, and at our depth, we will detect a small number of stars compared to the inner disk.  Thus, modeling the contribution of background galaxies to the CMDs is essential for measuring a reliable SFH.  \cite{2011ApJS..195...18R} shows in similar outer disk observations that removing background contaminates will drastically reduce the statistical uncertainties in the measurements.  Our outer most field 4 shows no evidence of upper main sequence or a even red giant branch.  Therefore we use this field as a model for background.\\
			\indent MATCH offers the option of specifying a CMD to include as a model for background contaminants, which we use across all fields.  This leaves us data for three of the four fields, but we find very little SF in field 3, telling us we would not have been sensitive of SF past at least 14.1 kpc.  This is to be expected, as a direct comparison of the field 3 and 4 CMDs shows that almost all the stars appear to be contaminants.\\ 
			
		\subsubsection{Uncertainties}\label{sec:error}
			Uncertainties due to photometric sampling are a source for random uncertainties, but low star counts represent the dominant term in the statistical uncertainties.  With limited stars, individual stars can have a significant impact on the outcome of our SFH.  Normal approaches of determining the random uncertainty i.e., calculating the probability density function or a Monte-Carlo (MC) algorithm, fail to retrieve the random error in cases of low SF \citep{2013ApJ...775...76D}.  With a low star count we expect to see low SF or none at all in the fits.  We then employ a Hybrid Monte-Carlo algorithm outlined in \citep{2013ApJ...775...76D} that retrieves the random errors even in cases with zero SF.  We invoke this algorithm within MATCH, which proceeds to sample the fitting solution space numerous times to derive the relative uncertainties for our SFH.\\
			\indent Sources for systematic errors would not only be from any limitations of the instruments used, e.g., UVIS vs ACS for redder stars, but also from the isochrones used.  The treatment of the systematic uncertainties is outlined in \citet{2012ApJ...751...60D}.  We, however, refrain from determining the systematics because we are interested in a comparison of the young stellar mass surface density of each field.\\
			\indent To get a reliable measurement in the stellar mass surface density of young stars, we consider the total SF for the past 200 Myrs, which is done by summing all the SF across the interested time from the best fit.  In order to determine the absolute errors we combine the first bins making up the first 200~Myrs, after fitting, with MATCH giving us our upper and lower bounds directly.

\section{Results}\label{sec:results}

\begin{figure}[b]
		\centering
		\includegraphics[scale=0.6]{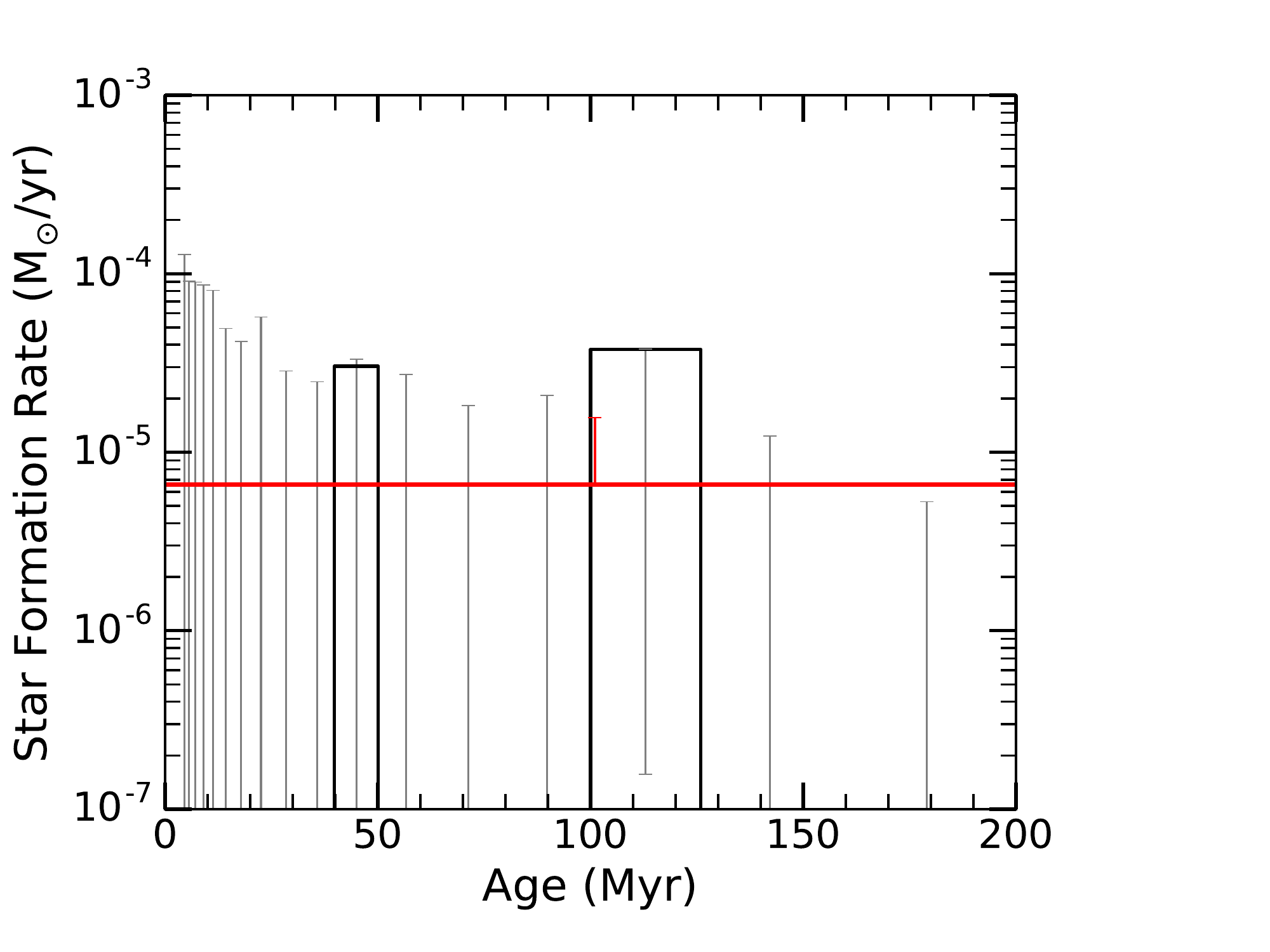}
		\caption{Plotted is an example SFH for field 2 spanning 200 Myrs from 6.6 to 8.3 log years with dex of 0.1 for a total of 17 bins.  On the vertical axis is the stellar mass computed for that particular bin.  Statistical uncertainties are shown in gray and derived from the MATCH algorithm utilizing a hybrid Monte-Carlo technique following \citet{2013ApJ...775...76D} (see Section \ref{sec:error}).  The errors are notably high in each bin due to the covariance of the neighboring bins.  For example, when there is a negative uncertainty the neighboring bin will have a high positive error because MATCH is not sensitive to which bin the star formation should be.  For comparison, we include, in red, the average SFR for the entire time range (which has relatively small uncertainties using our technique outlined in \ref{sec:error}).}
		\label{fig:SFH}
	\end{figure}

\begin{figure}[b]
		\centering
		\includegraphics[scale=0.6]{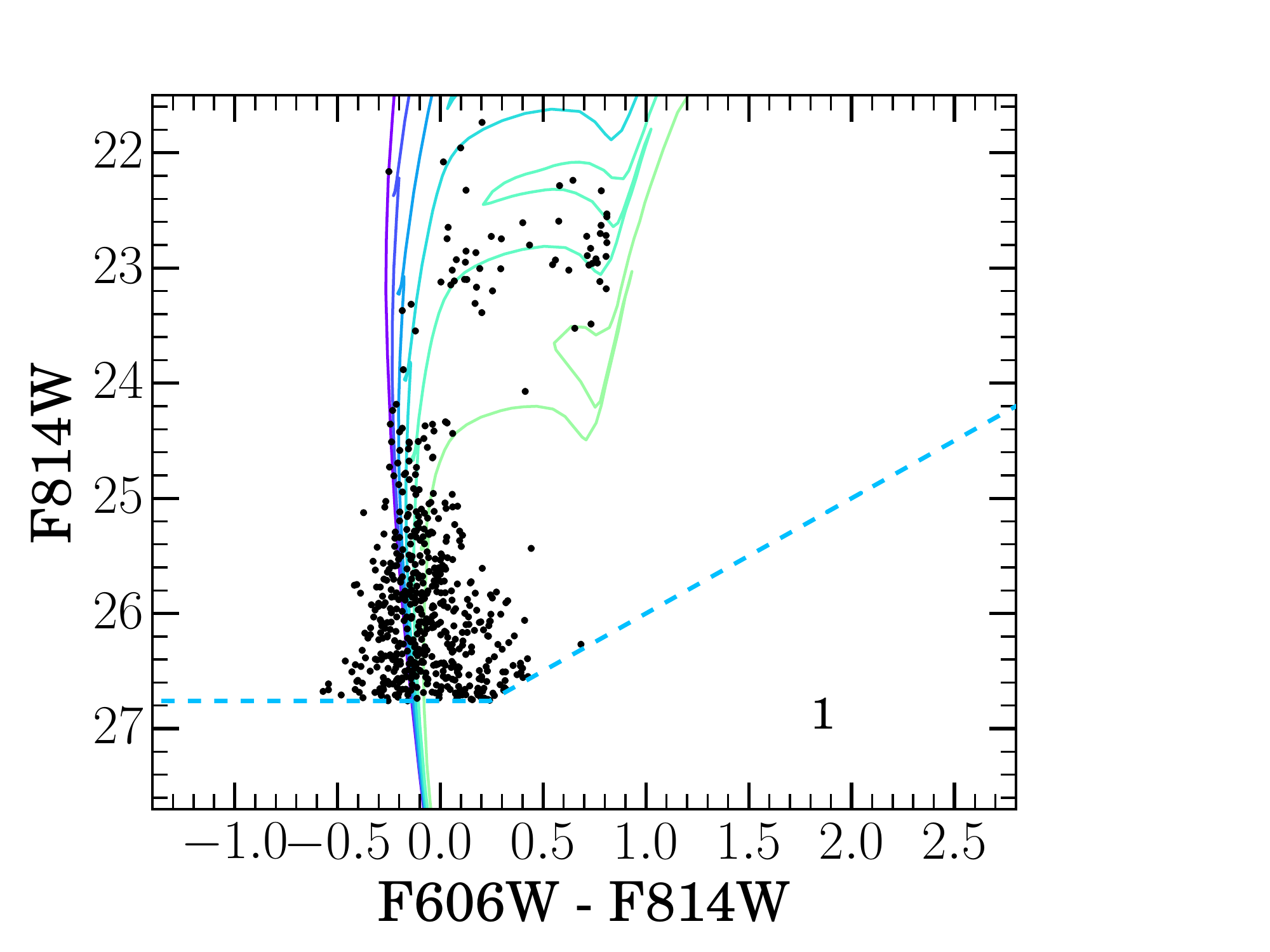}
		\caption{Best fits are run back through MATCH retrieving, randomly, the modeled stars for the past 200~Myrs.  These stars are represented through a similar CMD to Figure \ref{fig:cmd} where the blue dashed line is the chosen completeness limit and there are isochrones ranging from 7-8.3 log years.  Plotted as black points are the 576 stars that the MATCH emulation produces for field 1.  In similar fashion, MATCH produces 24 and 0 stars respectively in fields 2 and 3.}
		\label{fig:fake}
	\end{figure}

\begin{figure}[b]
		\centering
		\includegraphics[scale=0.6]{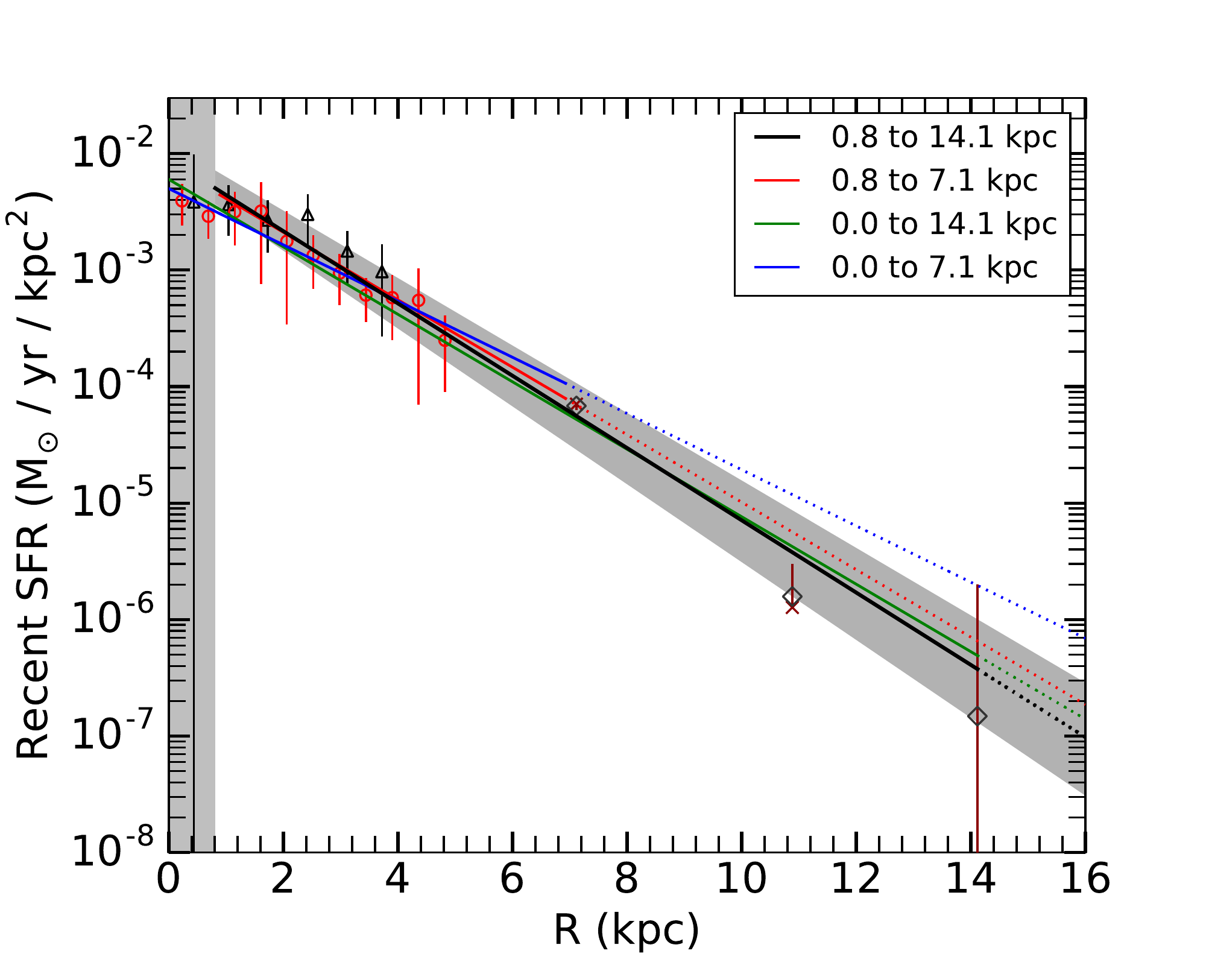}
		\caption{Plotted is the young mass stellar density for stars formed in the past 200 Myr as a function of galactocentric radius.  Our data are represented by dark red X's while the inner HST and GALEX-Spitzer $24\mu$ data are represented by triangles and circles respectively.  In addition, black diamonds represent the measurements derived when testing a more aggressive cut of the redder population.  We plot four different fits with the main one of interested designated as black bounded by a shaded region illustrating the area covered by the errors of this line.  Fits within the vertical shaded region include the three points within, and dotted lines are seen as extrapolations.  The last point is represented only by an upper bound.  See Section \ref{sec:results} for further discussion.}
		\label{fig:results}
	\end{figure}

		Using the best fit, we use the measured SFH to calculate the young stellar surface mass density of stars with ages $<$ 200 Myrs for each field.  We start with our SFHs with the full resolution of time specified in Section \ref{sec:match}.  Figure \ref{fig:SFH} is an example SFH of field 2 for the past 200 Myrs with the Hybrid MC errors included.  The relative errors for each bin are large due to high covariance from bin to bin.  The covariance can be seen especially in places where there is high negative relative uncertainty and the next bin over has a large positive uncertainty.  This is due to fitting giving the best result in only one of the bins versus other acceptable fits with the SF in the neighboring bin, showing that the fit is not sensitive to any individual time bin.  Thus, we report a single average SFR over the entire time range plotted as a red line in Figure 3.\\
		\indent We run the best fits back through MATCH to generate the number of stars MATCH fitted for our interested time frame of $<$~200~Myrs.  Plotted in Figure \ref{fig:fake} are those stars fitted for field 1 showing MATCH modeled 576 star for field 1; fields 2 and 3 have modeled counts of 24 and 0 stars, respectively, over 200~Myrs. \\
		\indent Figure \ref{fig:results}, shows the stellar mass surface density as a function of galactocentric radius in for stars younger than 200 Myr.  Previously measured inner points derived from HST imaging \citep{2010ApJ...712..858G} are represented by triangles, and azimuthally averaged points using GALEX-Spitzer $24\mu$ data \citepalias{2013ApJ...765..120W} are represented by circles, which help to extend the measurements out within the inner radii.  Our main points (dark-red X's) show significant detected stellar mass in 2 of the three fields with an upper limit in the last.  Thus, the outermost field, field 4, is unlikely to contain enough young stars to measure, and is a reliable background sample. Presented as black diamonds are the measurements from the aggressive exclusion of stars (see red dotted line in Figure \ref{fig:cmd}) showing that our results are not sensitive to this portion of the CMD.\\
		\indent Overplotted are four exponential fits using a $\chi^2$ goodness of fit with the vertical shaded region representing an area where two of the fits do not include the points enclosed, and any extrapolation is in the form of a dotted line.  We present the statistics of the fits in Table~\ref{tab:scale} that include the range of the fit, the $\chi^2$ per degrees of freedom, the normalization, and the scale length all in respective order.  The blue line in Figure~\ref{fig:results} represents a fit without our data points, and is characterized with a larger scale length.  Comparatively, the red line, which leaves out the three inner most points, is consistent with our points, that are not fitted to the red line.  This indicates a lack of recent star formation in the central regions of NGC~300.  The main fit, indicated as the black line, excludes the inner points to avoid this bias, though we give the full fit in green for comparison.  However, the measurements have enough leverage that the scale lengths measured are consistent with or without the central points.  The shaded region following the black line represents the area contained within the errors, illustrating the consistency in the red and green fits.  Our best derived scale length for the young stellar mass is $1.4\pm 0.1$ kpc.\\
		\indent We find no break in the young stellar disk of NGC~300 out to at least $\sim$10.9 kpc, or $\sim$8 scale lengths.  This is consistent with \citet{2005ApJ...629..239B}, which finds there to be no break in the total surface brightness profile out to $\sim$14 kpc, or 10 scale lengths.  This measurement is also consistent with that of \citet{2010ApJ...712..858G} whom find 1.3$\pm$0.1 kpc in the inner disk.  Thus, the young disk appears essentially undisturbed, similar to NGC~2403 \citepalias{2013ApJ...765..120W}. \\
		
	\section{Discussion: Environment and Outer Disk Evolution} 
			Now that we have determined the undisturbed nature of the young component in the outer NGC~300 disk, we use it to inform on the effects of environment on outer disk evolution through comparisons of the similar nearby disk galaxies M33 and NGC~2403.  By reporting on the gas disk as traced by the HI and oxygen abundances and the stellar disk through surface brightness, metallicity gradient, cosmic evolution scale length, and young stellar density we can comment on the effects of the environment across these three galaxies.\\
			\indent Gas disks can enlighten us to recent interactions within these three galaxies, we can effectively trace the shape through HI and gas-phase metallicity measurements.  Observations show the HI disk of M33 is warped past the optical disk \citep{1989AJ.....97..390C}, and a past interaction with M31 is the likely cause \citep{2009ApJ...703.1486P}.  In plain contrast, the HI disk of NGC~2403 is pristine and unbroken out to at least $\sim$15 kpc \citep{2002AJ....123.3124F, 2008AJ....136.2648D}.  In this regard, NGC~300 may appear similar to M33 with a warp starting at $\sim$6 kpc \citep{1990AJ....100.1468P, 2011MNRAS.416..509H}.  In addition, the warp in NGC~300 could be caused by the smaller neighbor NGC~55 which also shows evidence of a similar warp \citep{2011MNRAS.410.2217W, 2013MNRAS.434.3511W}; however, the case of NGC~300 is less clear than that of M33. For example, gas streams lying along the axis between M33 and M31 are indicative of a past interaction \citep{2009ApJ...703.1486P,2013ApJ...763....4L}, whereas NGC~300 and NGC~55 have no strong commonalities.\\
			\indent While the shape of the HI gas disk provides clues on the dynamical history, the abundance gradient of the ionized regions of the gas can also inform us about evolution.  For example, the gas-phase oxygen metallicity in simulations is shown to flatten in the presence of mergers on time scales of less than $\sim$ 1 Gyr \citep{2010ApJ...710L.156R, 2012ApJ...746..108T,2015ApJ...806..267Z}.  This suggests oxygen in radial distributed HII regions is a potential tracer of interaction.  Indeed, the abundance gradients in all non-interacting galaxies appears to be similar regardless of mass, morphological type, surface brightness, etc. \citep{2014A&A...563A..49S}.  A study of known perturbed galaxies versus a control group of isolated galaxies yielded $-0.23\pm0.03$ dex/$R_{25}$ and $-0.57\pm0.05$ dex/$R_{25}$ oxygen gradients, respectively, showing a clear difference in the two quantities \citep{2010ApJ...723.1255R}.  Thus, interactions appear to flatten the abundance gradient of the ionized gas, showing perturbations could cause breaks in SF if they already appear to effect the star forming gas.  In this regard, NGC~2403 and NGC~300 \citep[with gradients of -0.524$\pm$0.043, -0.519$\pm$0.040 dex/$R_{25}$ respectively,][]{2014AJ....147..131P} are both similar to the average value for non-interacting galaxies of $-0.57\pm0.05$ dex/$R_{25}$, and even when excluded from the control sample, the average changes to -0.55 dex/$R_{25}$ showing no skew in the results.  Conversely, M33 \cite[-0.359$\pm$0.041 dex/$R_{25}$,][]{2014AJ....147..131P} is clearly flatter than the others, which is consistent with a past interaction.  Thus, while the gas disk shape of NGC~300 may suggest recent interaction, the abundance gradient suggests no significant interaction.  This discrepancy leaves open the possibility that another process has warped the HI disk.\\
			\indent With somewhat conflicting information about interactions coming from the gas disks, we can turn to the details of the stellar disk for more clues.  The stellar disk as traced by brightness and stellar metallicity can also be influenced by gravitational interactions.  For example, the surface brightness profile of the disk can be broken by such interactions.  Similar to the oxygen abundance gradient, the stellar surface brightness profile of M33 is different from those of NGC~300 and NGC~2403.  M33 is known to have a break in the stellar surface brightness profile while those of NGC~2403 and NGC~300 are unbroken as far as can be measured \citep{2005ApJ...629..239B, 2007iuse.book..239F, 2012MNRAS.419.1489B}.  This is in line with the conclusions drawn of the gas-phase metallicity gradients.  This suggests that M33 is the only one to have likely encountered a gravitational interaction strong enough to change the stellar disk.\\
			\indent We can further infer on the evolutionary past of the stellar profile by observing how the scale length has transformed through cosmic time.  Characteristically, NGC~2403 and NGC~300 have seen little temporal evolution in the stellar disk scale length \citetext{\citetalias{2013ApJ...765..120W}; \citealt{2015arXiv151004768K}}, suggesting a quiet past in SF.  Conversely, M33 is measured to have had a scale length growth by a factor of $\sim$2 \citep{2009ApJ...695L..15W}.  This is accompanied by a break in the M33 stellar profile designating a shift from ``inside-out" growth to ``outside-in" past the break at $\sim$9 kpc \citetext{see \citealt{2011MNRAS.410..504B} for causation scenarios}.  Here, again, M33 is different  from NGC~300 and NGC~2403, potentially suggesting stronger gradients in less isolated systems.\\ 
			\indent The young stellar density profiles also give an indication for past interactions, recent interaction would likely cause a break in this profile or differences between this profile and the total surface brightness profile.  In this regard, NGC~300 is again similar only to NGC~2403.  See Figure \ref{fig:results} and \citetalias{2013ApJ...765..120W} Figure 10 for analogous plots of M33 and NGC~2403.  The young stellar surface density profile in M33 is much more centrally concentrated than the other two.  On the other hand, like NGC~300, NGC~2403 has a young star profile similar to its total surface brightness profile and NGC~2403.  This is again consistent with NGC~300 not having undergone significant interactions, and the presence of NGC~55 has had little influence over the evolution of NGC~300.\\
			\indent By separating the young stellar disk of NGC~300 we can also put new constraints on the total structure of NGC~300.  In NGC~300, the old stellar metallicity gradient flattens in the outer disk \citep{2009ApJ...697..361V}, which could indicate a transition from the stellar disk to a halo component, or internal scattering of stars to the outer disk \citep{2008ApJ...675L..65R}.  In addition, the total surface brightness cannot distinguish between an unbroken disk and a broken disk with a halo mimicking a continuous profile.  We find the stellar disk in NGC~300 has no break, suggesting the stellar disk continues despite a flattening in the metallicity.  This profile leaves little room for a halo component.  Thus it appears that secular mechanisms within the disk, such as radial migration, are more likely the cause of the metallicity flattening in NGC~300. \\
			\indent Finally, other than the HI warp, all other evidence appears consistent with no significant environmental influences on the outer disk of NGC~300.  Therefore, while the HI warp in M33 could likely be due to gravitational perturbation, the HI warp in NGC~300 is more likely attributable to infalling gas than to gravitational perturbations.

\section{Conclusions}\label{sec:conclusion}
	We measured resolved stellar photometry on four HST fields in the outer disk of NGC~300.  We determined the recent stellar mass surface density from CMD fitting (using the package {MATCH}) out to $\sim$8 scale lengths with an upper limit out to $\sim$10 scale lengths.  We find that the young population in NGC~300 is not centrally concentrated in the disk nor is there a break in the young stellar mass surface density, indicating NGC~300 has an undisturbed star forming outer disk.  This young stellar mass surface profile is consistent with the unbroken total surface brightness profile.\\ 
	\indent With this new knowledge of the young outer disk component of NGC~300, we revisited comparisons of the properties of the NGC~300 disk with those of 2 similar disk galaxies NGC~2403 and M33.  While M33 appears consistent with having undergone gravitational interactions, NGC~2403 and NGC~300 do not.  In this regard, the HI warp in NGC~300 is inconsistent with a gravitational interaction scenario and instead more likely attributed to infalling gas. Furthermore, the similarity of the young disk and total stellar profiles in NGC~300 suggest there is no significant halo component.  We conclude that NGC~300 and NGC~2403 have then seen no significant interactions in the past compared to the apparent evolution of M33. \\

	\indent Support for this work was provided by NASA through the grant GO-13461 from the Space Telescope Science Institute, which is operated by the Association of Universities for Research in Astronomy, Incorporated, under NASA contract NAS5-26555. We acknowledge the anonymous referee’s attention to the details of our writing style and presentation.\\


\newpage
\begin{deluxetable}{lllccccc}
	\tablewidth{0pt}
	\tabletypesize{\scriptsize}
	\tablecaption{Summary of Photometric Measurements\label{tab:fields}}
	\tablehead{ \colhead{Target (Field)} & \colhead{Camera} & \colhead{Filter} & \colhead{Exp. (s)} & \colhead{R (kpc)} & \colhead{Area (kpc$^2$)} & \colhead{Stars (Observed/Fit)} & \colhead{$m_{50\%}$} }
	\startdata
	NGC-300-OUTER-1U (1) & UVIS & F606W & 1000 & 7.1 & 3.33 & 4477/4459 & 27.62 \\
	NGC-300-OUTER-1U (1) & UVIS & F814W & 1500 & 7.1 & 3.33 & 4477/4459 & 26.81 \\
	NGC-300-OUTER-1 (2) & ACS & F606W & 894 & 10.9 & 5.19 & 960/950 & 27.72 \\ 
	NGC-300-OUTER-1 (2) & ACS & F814W & 1354 & 10.9 & 5.19 & 960/950 & 27.08\\ 
	NGC-300-OUTER-2U (3) & UVIS & F606W & 1000 & 14.1 & 3.33 & 269/267 & 27.63\\
	NGC-300-OUTER-2U (3) & UVIS & F814W & 1500 & 14.1 & 3.33 & 269/267 & 26.77\\
	NGC-300-OUTER-2 (4) & ACS & F606W & 894 & 17.7 & 5.19 & ... & 27.71\\
	NGC-300-OUTER-2 (4) & ACS & F814W & 1354 & 17.7 & 5.19 & ... & 27.10\\	
	\enddata
\end{deluxetable}	

\begin{deluxetable}{ccccc}
	\tablewidth{0pt}
	\tablecaption{Exponential Fits\label{tab:scale}}
	\tablehead{ \colhead{Start (kpc)} & \colhead{End (kpc)} & \colhead{$\chi^2/\nu$} & \colhead{Normalization} & \colhead{Scale Length (kpc)} }
	\startdata
	0.8 & 14.0 & 0.52 & $9^{+3}_{-2}\times 10^{-3}$ & $1.4_{-0.1}^{+0.1}$\\
	0.8 & 7.1 & 0.22 & $ 8^{+5}_{-4}\times 10^{-3}$ & $1.5^{+0.5}_{-0.3}$\\
	0.0 & 14.1 & 0.69 & $ 6^{+1}_{-1}\times 10^{-3}$ & $1.5^{+0.1}_{-0.1}$\\
	0.0 & 7.1 & 0.28 & $ 5^{+2}_{-1}\times 10^{-3}$ & $1.8^{+0.2}_{-0.3}$\\ 	
	\enddata
\end{deluxetable}

\end{document}